\documentstyle[12pt,makeidx,epsfig,rotating]{article}
\newcommand{\bp}{\begin{picture}}
\newcommand{\ep}{\end{picture}}
\include{def}

\begin{document}
\begin{titlepage}
\par 
\begin{center}
{\large EUROPEAN ORGANIZATION FOR NUCLEAR RESEARCH \\
\bigskip
\hfill 
\hfill \normalsize CERN-EP/99-02} \\
{\hfill  \normalsize January 4, 1999} \\
\vfill 
{\Large \bf Distinction of atmospheric $\nu_\mu-\nu_\tau$ and $\nu_\mu-\nu_s$
 oscillations using short or intermediate baseline experiments}
\vspace*{0.3cm}
\vfill 
\vspace*{0.4cm}
\normalsize 
\centerline{A. Geiser 
\footnote{CERN, EP Division, CH-1211 Gen\`eve 23,\quad Tel. +41-22-7678564, 
\quad Fax +41-22-7679070,\\ \qquad e-mail Achim.Geiser@cern.ch}
\footnote{now at Lehrst. f. Exp. Physik IV, Dortmund University,
Dortmund, Germany}}
\vfill 
\begin{abstract}
\indent 
\textwidth 8.0cm 
\hoffset 5.0cm
The current case for atmospheric $\nu_\mu$ oscillations into active or sterile
neutrinos is reviewed. It is argued that neither the study of neutral 
current events at Super-Kamiokande, nor the information obtained from future 
long baseline experiments might be sufficient to unambigously decide between
these two scenarios. However, a combination of these results with the results 
from future short or intermediate baseline $\tau$ appearance experiments 
would clearly resolve most of the remaining ambiguities. 
This conclusion does not strongly depend on whether the results from LSND 
will be confirmed or not. In the case that LSND would be confirmed, a negative 
result in such a short or intermediate baseline experiment would also 
unambigously exclude the 
interpretation of LSND as indirect $\nu_\mu-\nu_\tau-\nu_e$ oscillations. 
\end{abstract} 

\vspace*{1cm}
{\sl to be published in Eur. Phys. J. C} 


\bigskip 
keywords: atmospheric, neutrino, nutau, sterile, oscillations, short, 
intermediate, baseline   
\end{center} \vfill 


\end{titlepage}

\hoffset -0.7in 
\textwidth 6.0in 
\textheight 9.0in 
\normalsize 
\pagenumbering{arabic}
%
%

\section{Introduction}

In the wake of the recent evidence for atmospheric $\nu_\mu$ oscillations by 
Super-Kamiokande \cite{SuperK}, one of the crucial issues is to 
clarify whether the observed effect is due to
$\nu_\mu-\nu_\tau$, $\nu_\mu-\nu_e$, or $\nu_\mu-\nu_s$ (sterile) 
oscillations. The pure $\nu_\mu-\nu_e$ case is strongly disfavoured 
\cite{numunueex} by CHOOZ~\cite{CHOOZ} and Super-Kamiokande \cite{SuperK}, 
while 
pure $\nu_\mu-\nu_\tau$ and $\nu_\mu-\nu_s$ oscillations are equally
possible~\cite{taustecomp}. Within the restrictions imposed by 
CHOOZ, complicated
mixtures of the three cases can also not be excluded experimentally
\cite{numunue}\cite{numunuenus}.
In fact, most models favouring the sterile neutrino interpretation 
\cite{Pseudo}\cite{Koide}\cite{Mirror}\cite{stGUT} do suggest some (small) contribution 
from standard flavour
oscillations. 
Due to the fundamental implications of the existence 
of a sterile neutrino for new physics, we will assume that a separation
power of 3 to 5 $\sigma$ is desired to rule out the mainly sterile oscillation 
scenario,
and that a significance of at least 5 $\sigma$ is required to establish it. 

The paper is organized as follows: first, we will review the current
experimental status of the active versus sterile
oscillation hypothesis. We will then give some arguments why it is likely that
future improvements on these measurements by atmospheric and long baseline
neutrino experiments will leave important loopholes in the confirmation
of either of the two hypotheses, and how a modest admixture of 
$\nu_\mu-\nu_\tau$ to mainly $\nu_\mu-\nu_s$ oscillations can mimic the
pure $\nu_\mu-\nu_\tau$ oscillation case. Finally, we will show how these 
loopholes
can be closed by using the (positive or negative) results from
Short or Intermediate Baseline $\tau$ Appearance (SIBTA) experiments
\cite{TOSCA}\cite{COSMOS}\cite{JURA}.

Furthermore, we will argue that 
this conclusion does not depend strongly on whether the LSND observation of 
$\nu_\mu-\nu_e$ oscillations \cite{LSND} will be confirmed or not. 
If LSND would be confirmed, a negative result in a SIBTA experiment 
would also unambigously exclude the
interpretation of LSND as indirect $\nu_\mu-\nu_\tau-\nu_e$ oscillations,
therefore ruling out all oscillation scenarios invoking this option
\cite{Babu}. This includes essentially all models trying to reconcile
LSND with the atmospheric and solar neutrino result in the standard 
3-neutrino framework \cite{all3}.
  
\section{Discussion of Super-Kamiokande indications}

Results of neutrino oscillation experiments are often expressed in terms of 
an effective two flavour oscillation scheme with a mixing angle 
$\sin^2 2\theta$ between the two flavours and a mass difference $\delta m^2$
between the two relevant mass eigenstates.
The most popular interpretation of the Super-Kamiokande results is to invoke
maximal or close to maximal $\nu_\mu-\nu_\tau$ oscillations with a 
$\delta m^2$ in the $10^{-2} - 10^{-3}$ eV$^2$ range. Clearly, if this 
interpretation is correct, the low $\delta m^2$ excludes any observation of a 
$\nu_\mu-\nu_\tau$ oscillation signal in current \cite{NOMAD}\cite{CHORUS}
and future \cite{TOSCA} short baseline 
experiments. However, a significant $\nu_\mu-\nu_e$ contribution to
the angular dependence of the atmospheric neutrino result is not completely 
excluded, and even suggested by
some of the models \cite{all3} trying to reconcile the 
atmospheric neutrino anomaly with LSND. In this case
the constraint on $\delta m^2_{\mu\tau}$ could be 
considerably diluted, and $\tau$ appearance signals in short or intermediate 
baseline experiments could be possible.

Alternatively, several theoretical models suggest the 
interpretation of the Super-Ka\-mio\-kan\-de results as $\nu_\mu-\nu_s$ oscillations.
In this case, the sterile neutrino could either be the right-handed (sterile) 
partner of the left-handed (active) muon neutrino (or its antiparticle), 
leading to maximal $\nu - \bar\nu$ oscillations analogous to 
$K^0-\bar{K^0}$ oscillations \cite{Pseudo}\cite{Koide}, 
or the light remnant of very 
massive neutrinos in GUT extensions of the standard model usually
involving extra neutrino multiplets \cite{Mirror}\cite{stGUT}. 
In most of these models the observation of $\nu_\mu-\nu_s$ 
oscillations could be further complicated by non-negligible
admixtures of $\nu_\mu-\nu_\tau$ or $\nu_\mu-\nu_e$ oscillations.

Since, due to the high $\tau$ mass threshold, neither $\nu_\tau$ nor $\nu_s$
produce a visible charged current (CC) signal in Super-Kamiokande, 
the two cases are experimentally almost indistinguishable in the standard 
Super-Kamiokande analysis
of e-like and $\mu$-like events \cite{SuperK}\cite{taustecomp}. 
However, $\nu_\tau$'s do produce neutral
current (NC) interactions while sterile neutrinos do not. This could lead to a
visible distinction in two kinds of measurements: In the $\nu_\mu-\nu_s$
case the up/down asymmetry observed in the CC
sample should also be present for NC events, while no
NC up/down asymmetry should occur in the $\nu_\mu-\nu_\tau$
(or $\nu_\mu-\nu_e$) case \cite{NCasym}. This effect also yields 
differences for the up/down asymmetry in inclusive event samples 
\cite{multiring}. Furthermore, the NC suppression in 
the $\nu_\mu-\nu_s$ case modifies the predicted NC/CC ratios \cite{NCCC}.

Unfortunately a clean NC/CC separation is experimentally difficult, and the
expected effects are diluted by the (supposedly) unaltered
contribution from $\nu_e$ NC events. 
The cleanest way to identify NC events in Super-Kamiokande is to require a
single $\pi^0$ from the process $\nu + N \to \nu + N + \pi^0$, with N being
either a neutron or a proton below \v{C}erenkov threshold. The $\pi^0$ is
detected via its decay into two photons which convert and yield two 
electron-like (double) rings whose invariant mass is consistent with the 
$\pi^0$ mass.
This procedure reduces statistics by so much that 
currently no significant measurement of the up/down asymmetry can be obtained
\cite{SuperKNOW98}.
The ratio of two ring ($\pi^0$-like, NC) to single ring (e-like, CC) events 
compared to the 
prediction for the no oscillation case is measured to be 
\cite{SuperKNOW98}\footnote{The value of 0.94 quoted e.g. in \cite{Takayama}
does not yet include background subtraction.}
$${{(\pi^0/e)_{data}}\over{(\pi^0/e)_{pred}}} = 0.88 \pm 0.08_{stat} \pm 0.19_{sys}\quad (1)$$
where the systematic error is dominated by the 
poorly known single $\pi^0$ cross section, and the statistical error is based 
on 535 days (2 years) of running. Assuming an initial 
$\nu_\mu / \nu_e$ ratio of 2.0, using the measured average $\nu_\mu / \nu_e$ 
ratio suppression of 0.63 \cite{SuperK}, and assuming no background, the ratio
in eq. (1)
is expected to be 1.00 for $\nu_\mu-\nu_\tau$ oscillations (neither
of the two contributions is affected), 0.75 for $\nu_\mu-\nu_s$
oscillations (the $\pi^0$ contribution is suppressed), 0.75 for  
$\nu_\mu-\nu_e$ oscillations (the $e$ contribution is enhanced), and 0.94
for mainly $\nu_\mu-\nu_\tau$ oscillations with a 10\% $\nu_\mu-\nu_e$ 
contribution. If there is significant background contamination these 
differences will be further reduced.

In order to disentangle  $\nu_\mu-\nu_\tau$ from
$\nu_\mu-\nu_s$ at the 3 $\sigma$ level, the combined statistical 
and systematic error must therefore be reduced to 8\% or less. If a 10\%
$\nu_\mu-\nu_e$ contribution is allowed, the required maximal uncertainty
is reduced to 6\%.
Arbitrarily assuming 2140 days (8 years) of running, the statistical error 
will be about 4\%. In order not to exceed the 6\% total error this implies a 
systematic uncertainty of the order of 5\% or less.
Even with the planned measurement of the $\pi^0$ production cross section
in the near detector of the K2K experiment \cite{K2K} this seems to be 
very hard to achieve.   
We therefore conclude that this measurement will probably yield a useful 
indication, but is unlikely to firmly establish one 
of the two options.

Similar arguments hold for the $\pi^0$ up/down asymmetry.
From a simple extrapolation of existing data \cite{SuperKNOW98}, it looks 
unlikely that this method will distinguish the two cases by more than about 
2~standard deviations.

The possibility to separate $\nu_\tau$ and $\nu_s$ from the measurement of
the inclusive up/down asymmetry of multiring events, which also depends on
the suppression of the NC contribution, is discussed in ref. \cite{multiring}.
Here, the statistical and systematic errors are smaller, but the differences 
are also small, again yielding a potential effect of about 2 $\sigma$.
Furthermore, $\nu_\mu-\nu_s$ oscillations with a significant $\nu_\mu-\nu_e$ 
admixture could yield the same asymmetry as pure $\nu_\mu-\nu_\tau$
oscillations.

\section{Other atmospheric neutrino experiments}

None of the currently planned atmospheric neutrino experiments 
\cite{NOE}\cite{NICE}\cite{GSatm}\cite{ICARUS}
has a $\tau$ detection efficiency which is sufficiently large to see a 
significant pure charged current oscillation signal. 
In addition to the small cross section ($\tau$ mass
threshold), the unknown direction of the incoming neutrino makes a significant
kinematical analysis ``\`a la NOMAD'' \cite{NOMAD} impossible. Emulsion 
techniques ``\`a la CHORUS'' \cite{CHORUS} can not be used due to the 
inherently small target mass.

As in Super-Kamiokande, any efforts to distinguish between $\nu_s$ and 
$\nu_\tau$ must therefore focus on neutral current events or on 
inclusive event rates. In a detector
like NICE \cite{NICE} there is a small window at 
$\delta m^2 \sim {\rm few}~10^{-4}$ eV$^2$ where the oscillation pattern could 
actually be resolved, opening the possibility of
comparing the energy dependence of the CC and NC event rates.
In addition, such a low $\delta m^2$ would in itself be an indication 
for oscillations into active neutrinos, since the sterile case is 
somewhat disfavoured for such low $\delta m^2$ values due to earth matter 
effects which start to play a role \cite{taustecomp}.

Ref. \cite{GSatm} outlines an atmospheric neutrino detector concept 
which would 
allow to measure $\nu_\tau$ appearance via an enhancement of muon-less
events from $\tau$ decays at the highest accessible neutrino energies, 
where $\tau$ production is least suppressed. 
The nice feature of this concept is that $\nu_e$ events are effectively 
filtered out, therefore removing most of the $\nu_e$ background and the 
$\nu_\mu-\nu_\tau$/$\nu_\mu-\nu_e$ ambiguity. 
Combined with the NC suppression for $\nu_s$ this yields a 
$\nu_\mu-\nu_\tau$/$\nu_\mu-\nu_s$ separation of several standard deviations 
for $\delta m^2 \sim 5 \times 10^{-3}$ eV$^2$ or larger \cite{GSatm}.
For $\delta m^2 \sim 3 \times 10^{-3}$ the sensitivity is significantly 
reduced (oscillations of high energy $\nu_\mu$'s are suppressed), while
for even lower $\delta m^2$ the difference becomes marginal. 

For the Super-Kamiokande most favoured case of 
$\delta m^2 \sim 2 \times 10^{-3}$ eV$^2$ it is not 
clear at present whether the difficulties outlined above 
will allow to draw any firm conclusions concerning the distinction
of active and sterile neutrino oscillations.
Furthermore, no detector of the types discussed above has been endorsed 
or approved so far.

In principle $\nu_\mu-\nu_\tau$ and $\nu_\mu-\nu_s$ oscillations can also
be distinguished through the distortion of the momentum spectra of  
upward going muons due to matter effects \cite{Lipari}.
No conclusion has been reached so far from ongoing experiments \cite{MACRO}
due to large systematic errors. It is not clear at present whether these 
measurements can be improved sufficiently well to eventually allow a 
clear distinction. 
\vfill\goodbreak

\section{Accelerator neutrino experiments}

\begin{figure}
\epsfig{file=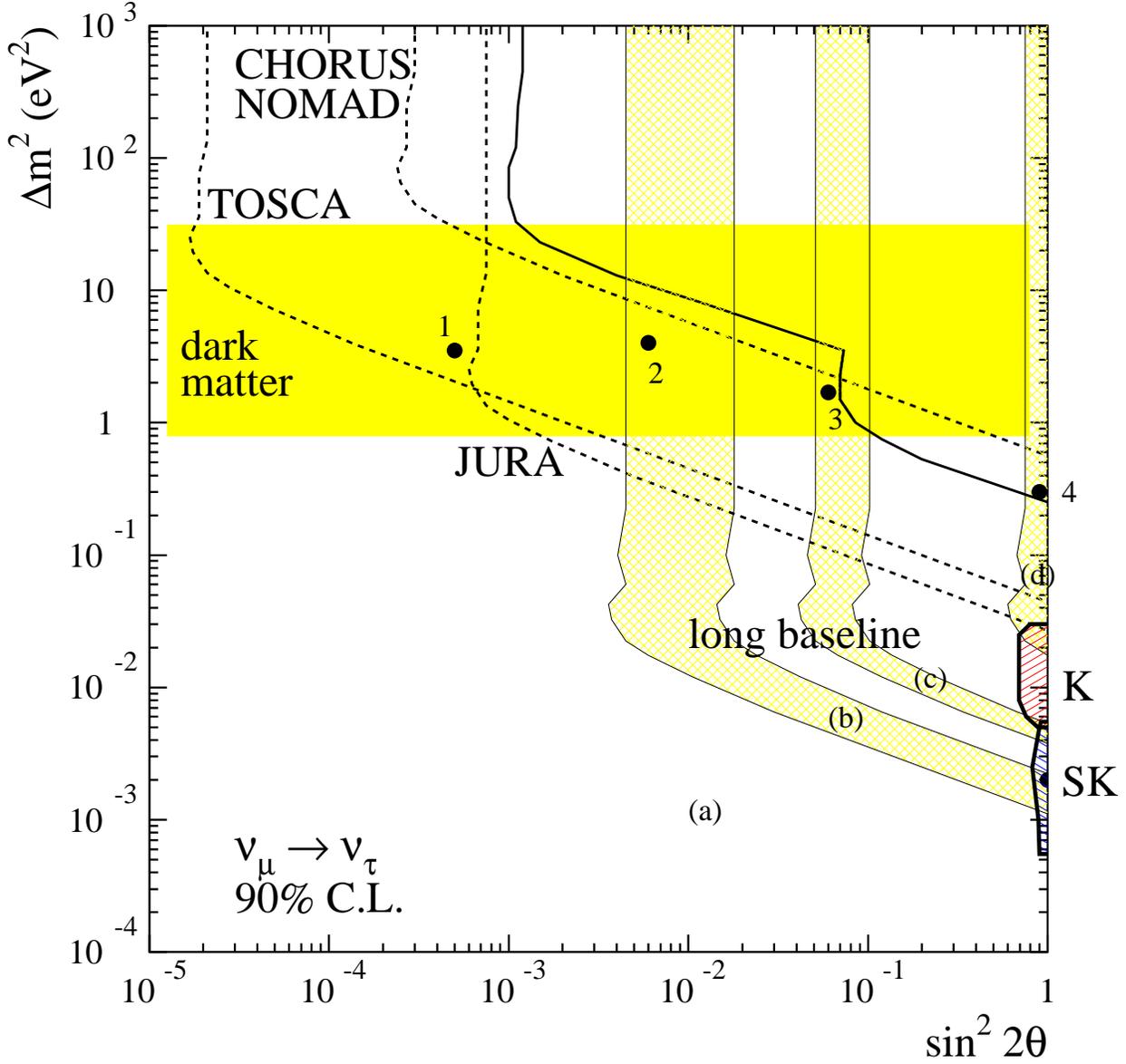,width=170mm}
\small
\caption{\footnotesize Parameter space for $\nu_\mu-\nu_\tau$ oscillations.
Indicated are the current combined limit from NOMAD \cite{NOMAD}, 
CHORUS \cite{CHORUS},
E531 \cite{E531} and CDHS \cite{CDHS} (thick continous line) as well as the
potential future limits from CHORUS/NOMAD, TOSCA \cite{TOSCA}, and a generic 
intermediate baseline experiment in the Jura \cite{JURA} (dashed lines).
The allowed regions of the Kamiokande (K) \cite{Kamatmo} and 
Super-Kamiokande (SK) \cite{SuperK} experiments, if interpreted as 
$\nu_\mu-\nu_\tau$ oscillations, are also
shown. The bands labeled (b), (c), and (d) correspond to allowed regions 
from different potential $\tau$ appearance results of a generic long baseline 
experiment as discussed in the text, while (a) stands for a null result.
The shaded area indicates the region favoured by mixed dark matter scenarios
\cite{DM} while the points indicate test points used for the discussion 
in the text. They correspond to a specific prediction of 
$\nu_\mu-\nu_\tau$ oscillations (in addition to maximal $\nu_\mu - \nu_s$
oscillations) of ref.  \cite{Koide} (1),
a generic test point compatible with most of the models in refs. 
\cite{Pseudo}\cite{Mirror}\cite{stGUT} (2), and two test points 
corresponding to indirect oscillation solutions for LSND \cite{Babu}
(3 and 4). In addition, point 3 can be considered a variant of point 2.}
\label{fig:numunutau}
\normalsize
\end{figure}

The best way to establish the $\nu_\mu-\nu_\tau$ interpretation of the 
atmospheric neutrino result obviously seems to be the detection
of $\nu_\tau$ appearance in long baseline accelerator experiments
\cite{ICARUS}\cite{OPERA}\cite{MINOS}\cite{AquaRICH}\cite{NOE}. Here, 
appearance could be established either directly through the observation of 
$\tau$ production, or indirectly via an enhancement of the 
NC/CC ratio, together with the non-observation of a large electron
appearance effect.
But, it turns out that a positive $\tau$ signal in e.g. ICARUS \cite{ICARUS}, 
OPERA \cite{OPERA}, or MINOS \cite{MINOS} would {\sl not} 
automatically prove the $\nu_\mu-\nu_\tau$ hypothesis for atmospheric
neutrinos, unless the corresponding $\delta m^2$ can be measured directly
from this signal. 
The basic argument is that, as illustrated in Fig. \ref{fig:numunutau}, even
a small $\nu_\mu-\nu_\tau$ contribution ($\sin^22\theta \sim {\rm few}~10^{-3}$ 
or larger) at large $\delta m^2$ (order eV$^2$) can yield a signal in long
baseline experiments (for both appearance and disappearance) that is similar 
in size or even larger than the 
expected effect from maximal $\nu_\mu-\nu_s$ oscillations, and can therefore
mimic maximal $\nu_\mu-\nu_\tau$ oscillations at small $\delta m^2$.
This argument is discussed in detail below.

\subsection{Motivation for mixed $\nu_\mu - \nu_s$ and 
$\nu_\mu - \nu_\tau$ scenarios}  

One of the most general arguments brought forward in favour of 
$\nu_\mu-\nu_s$ oscillations in the context of some models 
\cite{Pseudo}\cite{Mirror} 
is the possibility to combine a neutrino 
mass hierarchy and mixing pattern which are similar to the one 
observed in the quark and charged lepton sector (i.e. third generation
is heaviest, mixing angles are small) with maximal $\nu_\mu-\nu_s$
oscillations, while retaining a significant contribution to hot dark matter.
In mixed dark matter models, this would suggest a $\nu_\tau$ mass of order
eV. Similarity with the CKM matrix (\cite{CKM}) would suggest a 
$\sin^22\theta_{\mu\tau}$ of order $10^{-2}$ (e.g. test point 2 in Fig. 
\ref{fig:numunutau}).
Once sterile neutrinos are considered at all this 
is in some sense a ``natural'' possibility which should not be dismissed 
a priori, although it is by no means a necessity. 
Without requiring mass hierarchy, similar arguments for a possible 
$\nu_\mu \to \nu_\tau$ admixture apply for
the models of ref. \cite{stGUT}.
   
If LSND \cite{LSND} would be confirmed this argument would be strengthened 
further,
since the observation of $\nu_\mu-\nu_e$ oscillations in the range
$\delta m^2 \sim 0.2 - 2$~eV$^2$, combined with the hierarchy assumption,
would again suggest a $\nu_\tau$ mass in the eV range. 
In addition, it would make the sterile scenario more attractive, since the  
LSND result is hard to combine with the evidence from atmospheric and 
solar neutrinos in a three neutrino scheme. However, the argument given in the
previous paragraph
does {\sl not} vanish if LSND would be disproven. Instead, 
$\nu_\mu-\nu_e$ oscillations might play a significant role at lower
$\delta m^2$, either as an admixture to mainly $\nu_\mu-\nu_s$
atmospheric neutrino oscillations 
or as the solution of the solar neutrino problem \cite{solar}.   

Finally, let us digress and consider the case of indirect 
$\nu_\mu-\nu_\tau-\nu_e$ oscillations in LSND \cite{Babu}\cite{Interm}. 
Assuming mass hierarchy
and denoting the dominant mass components of $\nu_e$, $\nu_\mu$ and $\nu_\tau$
by $m_1$, $m_2$, and $m_3$ yields the mass relation $\delta m_{12}^2 << 
\delta m_{23}^2 \sim \delta m_{13}^2 \sim m_3^2$ known as one mass scale 
dominance \cite{OMSD}. 
In order for indirect oscillations to be detectable, $m_3^2$ must be in the 
range relevant for LSND, i.e. of order 1 eV$^2$, and the rate proportional to
$$ \sin^2 2\theta_{\rm LSND} = 4 |U_{e_3}U_{\mu_3}|^2 \quad\quad(2)$$
must be sufficiently large. Here, $U_{e_3}$ and $U_{\mu_3}$ are the relevant 
matrix elements of the general 3 neutrino mixing matrix.
Bounds on $U_{e_3}$ from Bugey \cite{Bugey} combined
with the LSND measurement of the rate in eq. (2) yield a prediction for 
$U_{\mu_3}$ as a function of $m_3^2$. 
All possible solutions are close to the limit from CDHS \cite{CDHS}.
Two (marginally) allowed
solutions, translated into $\sin^22\theta_{\mu\tau}$, are shown as test points
3 and 4 in Fig. \ref{fig:numunutau}. 

Test point 3 corresponds to a scenario very similar to that of test point 2,
where $\nu_\mu - \nu_\tau$ oscillations happen in addition to the 
$\nu_\mu - \nu_s$ oscillations responsible for the atmospheric neutrino
deficit. It is therefore also relevant outside of the LSND context.

Test point 4 represents the region of very large $\nu_\mu - \nu_\tau$ 
mixing used, among others,  for the Cardall/Fuller, Acker/Pakvasa 
and Thun/McKee schemes \cite{all3} without the need for sterile
neutrinos. 


\subsection{Interpretation of long baseline observations} 
For the purpose of this study we will consider close to maximal $\nu_\mu$ 
disappearance of atmospheric neutrinos to be an established fact, and 
anticipate that this will be 
confirmed\footnote{But note some caveats explained later in the text.} 
through a positive effect in the
ratio of the $\nu_\mu$ CC rate in near and far detectors of the long
baseline programme (K2K \cite{K2K}, MINOS \cite{MINOS}, NICE \cite{NICE}, ...). 
Whenever atmospheric $\nu_\mu-\nu_\tau$ or 
$\nu_\mu-\nu_s$ oscillations are mentioned in the next few paragraphs, it is 
understood that there could be a small (up to 10\%) contribution from 
$\nu_\mu-\nu_e$. A potentially even larger $\nu_e$ contribution is assumed 
to be measured and corrected for. 
Finally, Fig. \ref{fig:numunutau} implies that very 
similar conclusions can be obtained
from short and intermediate baseline experiments. We will therefore often
refer to a generic Short or Intermediate Baseline $\tau$ Appearance (SIBTA) 
experiment in the discussion.
\smallskip

{\bf case (a): Long baseline experiments do not observe $\nu_\tau$ appearance.}

A positive signal in a SIBTA experiment would then {\sl establish} that 
$\nu_\mu-\nu_\tau$ 
oscillations are outside of the range accessible to long baseline experiments,
but within the range relevant to mixed dark matter 
(e.g. test point 1 in Fig. \ref{fig:numunutau} for the short baseline case),
and {\sl force} the $\nu_\mu-\nu_s$ interpretation of the atmospheric
neutrino result.
The measured long baseline disappearance rate would yield a measurement 
of the $\delta m^2$ for $\nu_\mu-\nu_s$ oscillations, to be compared to the 
Super-Kamiokande result.

A negative result in a SIBTA experiment would {\sl exclude} any 
$\nu_\mu-\nu_\tau$ 
contribution to the long baseline signal from the cosmologically relevant 
range. It would therefore {\sl strongly suggest} the $\nu_\mu-\nu_\tau$ 
interpretation of the atmospheric neutrino oscillations at the low end 
of the Super-Kamiokande allowed $\delta m^2$ range, provided the observed long 
baseline disappearance signal (from a low energy beam) 
is consistent with this low $\delta m^2$ hypothesis.
\smallskip

{\bf case (b): Long baseline experiments observe a small $\nu_\tau$
appearance signal.} 

Since the appearance signal is small (e.g. a handful of events in the case 
of direct $\tau$ appearance), it would 
supposedly {\sl not} be possible to reliably measure the $\delta m^2$
from the energy distribution of the appearance signal alone.
Again, a positive signal in a SIBTA experiment would {\sl establish} 
that $\nu_\mu-\nu_\tau$
oscillations occur in the cosmologically relevant range
(e.g. test point 2 in Fig. \ref{fig:numunutau}),
and {\sl force} the $\nu_\mu-\nu_s$ interpretation of the atmospheric
neutrino result.
The combination of the SIBTA and long baseline results would unambigously 
fix the $\nu_\mu-\nu_\tau$ oscillation parameters.
If the $\nu_\mu-\nu_s$ oscillations would occur at the Super-Kamiokande 
best fit
point of $\delta m^2 \sim 2\times 10^{-3}$ eV$^2$ (SK), they would be partially
masked by the $\nu_\mu-\nu_\tau$ signal in the long baseline 
disappearance search. On the other hand, if the disappearance rate would turn 
out to be significantly larger than the appearance rate
($\delta m^2 > 2\times 10^{-3}$ eV$^2$), the former could be used to measure 
the $\delta m^2$ of the $\nu_\mu-\nu_s$ oscillations.

A negative SIBTA result would again {\sl exclude} any $\nu_\mu-\nu_\tau$
contribution to the long baseline signal from the cosmologically relevant
range, and therefore {\sl strongly suggest} the $\nu_\mu-\nu_\tau$
interpretation of the atmospheric neutrino oscillations with parameters close 
to the Super-Kamiokande best fit point. This can be cross-checked by requiring 
the appearance and disappearance rates to agree. The combination of the 
long baseline with the atmospheric neutrino results then yields a
precise measurement of the $\nu_\mu-\nu_\tau$ oscillation parameters.
\smallskip\goodbreak

{\bf case (c): Long baseline experiments observe a large $\nu_\tau$
appearance signal.} 

Given a large $\nu_\mu-\nu_\tau$ appearance signal, it might be possible 
to extract $\delta m^2$ from the energy distribution of the 
appearance signal alone, or at least to exclude a significant fraction of
the available parameter space. However the example of the LSND
experiment \cite{LSND} 
shows that this possibility can not be taken for granted: 
Despite a claimed excess of 50 events over a small background, LSND
is not able to differentiate the low and high $\delta m^2$ cases for
their $\nu_\mu-\nu_e$ oscillation signal. Definitely, a 
low/high $\delta m^2$ distinction at the
5 $\sigma$ level would not be obvious for the long baseline results.

The preference of the low $\delta m^2$ case could e.g. suggest the 
$\nu_\mu-\nu_\tau$ interpretation of the atmospheric neutrino deficit 
with a $\delta m^2$ in the Kamiokande/Super-Kamiokande overlap region 
($\sim 6 \times 10^{-2}$ eV$^2$). The requirement of {\sl not} observing a 
SIBTA signal would essentially {\sl eliminate} the whole large $\delta m^2$
range, including the regions suggested by dark matter and/or LSND 
(test point 3), and therefore strongly
enhance the confidence in the low $\delta m^2$ result.
Compatibility of the observed long baseline appearance and disappearance 
rates, although required, would {\sl not} yield any further separation
power, since a large $\tau$ appearance signal from e.g. test 
point 3 in
Fig. \ref{fig:numunutau} would completely mask the atmospheric oscillation 
effect, and also yield a consistent appearance/disappearance rate.  
  
The preference of the high $\delta m^2$ case would imply the interpretation
of the atmospheric neutrino anomaly as mainly $\nu_\mu-\nu_s$ oscillations.
If LSND would be confirmed, it could in addition 
imply the compatibility of the 
$\nu_\mu-\nu_\tau$ oscillation parameters with the indirect oscillation 
hypothesis for LSND (test point 3). Given the importance of such a result, 
the cross-check from the requirement of a positive SIBTA result would be 
absolutely {\sl essential}.
\smallskip
 
{\bf case (d): Long baseline experiments observe a close to maximal ($> 30\%$) 
$\nu_\tau$ appearance signal.}

Such a signal would be really spectacular, and inconsistent with the 
analysis of the Super-Kamiokande data in terms of two flavour oscillations.
Since it would point to a $\delta m^2$ larger than $10^{-2}$ eV$^2$ 
it would either
indicate a serious flaw in the Super-Kamiokande analysis, or require 
a three (or more) flavour scheme with significant contributions from two 
different $\delta m^2$, one high and one low. 

The energy distribution of the observed $\tau$ spectrum could already give
a serious indication of the relevant high $\delta m^2$. 
The rate of a positive SIBTA signal would however 
unambigously fix the $\delta m^2$ for $\nu_\mu-\nu_\tau$ oscillations,
and could decide whether it corresponds to the Cardall/Fuller or 
Thun/McKee solutions (test point 4) or to some lower $\delta m^2$ value.

A negative SIBTA  result would {\sl constrain} the $\delta m^2$ to
about $2 \times 10^{-2}$ eV$^2$.

In either case the up/down asymmetry pattern of the Super-Kamiokande data
suggesting a significant low $\delta m^2$ contribution would either 
have to be disproven, attributed to $\nu_\mu-\nu_e$ oscillations through
a corresponding observation in long baseline experiments, or 
attributed to a complicated mixture of $\nu_\mu-\nu_\tau$, 
$\nu_\mu-\nu_e$, and $\nu_\mu-\nu_s$ oscillations. 

\section{Compatibility with LSND}

As outlined in the cases discussed above, none of the scenarios discussed 
essentially requires the confirmation of LSND. On the other hand, a 
confirmation of LSND would significantly enhance the interest in the 
$\nu_\mu-\nu_\tau$/$\nu_\mu-\nu_s$ distinction, since the addition of one or 
more sterile neutrinos might then be the only solution to simultaneously 
describe all the data.
Also, the question whether the LSND signal is caused by direct or indirect 
oscillations becomes very relevant.
As can be seen from Fig. \ref{fig:numunutau}, a negative SIBTA signal 
would unambigously exclude the indirect oscillation possibility (test points
3 and 4), while a positive signal might allow it. 
A positive $\nu_\mu-\nu_\tau$ signal of kind (c) or (d) in the long baseline
experiments could confirm this scenario, while a signal of kind (b) or (a)
(no signal) would again exclude it.
Finally, reversing the argument, the observation or non-observation
of a signal in a SIBTA experiment before the LSND case is settled could, 
depending on the
context of the results of other experiments, indirectly contribute to the
LSND verification.

\section{Conclusion}

It has been shown that, whatever the outcome of future atmospheric and
long baseline neutrino 
experiments, the complementary information from a short or intermediate 
baseline $\tau$ appearance experiment could be crucial for the unambigous 
interpretation
of the long baseline results. Such an experiment is therefore needed 
to distinguish cleanly between the $\nu_\mu-\nu_\tau$ and $\nu_\mu-\nu_s$
interpretations of the atmospheric neutrino anomaly, and even a negative 
result is very relevant in this context. In the absence of this cross-check,
a small $\nu_\mu-\nu_\tau$ contribution at high $\delta m^2$ could mask
the long baseline $\nu_\mu-\nu_s$ signal, therefore yielding a false 
confirmation of the interpretation of the atmospheric neutrino anomaly in 
terms of $\nu_\mu-\nu_\tau$ oscillations. 
This conclusion does not depend on whether LSND is confirmed or not, 
although a confirmation of LSND would enhance the interest in 
resolving this ambiguity.  
If LSND would be confirmed, a negative signal in such a short or intermediate
baseline experiment would definitively rule out the indirect oscillation
interpretation of the LSND result. 

\section{Acknowledgements}

Constructive discussions with E. do Couto e Silva and J.J. Gomez-Cadenas are 
gratefully acknowledged. Special thanks go to the organizers of the NOW'98
neutrino workshop in Amsterdam, 6-9 September 1998, for stimulating the 
present work.


%
\end{document}